\documentclass{nature3}
\usepackage{graphicx}
\usepackage{float}
\usepackage{url}

\newcommand{\project}[1]{\textsl{#1}}

\newcommand{\Spitzer}{\project{Spitzer}}                            
  
\newcommand{\TESS}{\project{TESS}}  

\newcommand\aj{{Astron. J.}}
\newcommand\apj{{Astrophys. J.}}
\newcommand\apjl{{Astrophys. J. Lett.}}     
\newcommand\apjs{{Astrophys. J. Suppl.}}
\newcommand\aap{{Astron. Astrophys.}}
\newcommand\icarus{{Icarus}}
\newcommand\physrep{{Phys.~Rep.}}
\newcommand\mnras{{Mon. Not. R. Astron. Soc.}}
\newcommand\pasp{{PASP}}
\newcommand\nat{{Nature}}

\title{Absence of a thick atmosphere on the terrestrial exoplanet LHS~3844b}

\author{Laura Kreidberg$^{1,2}$, Daniel D.B. Koll$^{3, *}$, Caroline Morley$^{4, *}$, Renyu Hu$^{5,6, *}$,   Laura Schaefer$^{7}$, Drake Deming$^{8}$, Kevin B. Stevenson$^{9}$, Jason Dittmann$^{3,10}$, Andrew Vanderburg$^{4,11}$, David Berardo$^{12}$, Xueying Guo$^{12}$, Keivan Stassun$^{13}$,   Ian Crossfield$^{12}$, David Charbonneau$^{1}$,  David W. Latham$^{1}$, Abraham Loeb$^1$, George Ricker$^{12}$, Sara Seager$^{3,12,14}$, Roland Vanderspek$^{12}$}

\begin{document}

\maketitle

\begin{affiliations}
\item Center for Astrophysics $|$  Harvard \& Smithsonian, Cambridge, MA 02138
 \item Junior Fellow, Harvard Society of Fellows 
 \item Department of Earth, Atmospheric and Planetary Sciences, Massachusetts Institute of Technology, Cambridge, MA 20139
 \item Department of Astronomy, The University of Texas at Austin, Austin, TX 78712
\item Jet Propulsion Laboratory, California Institute of Technology, Pasadena, CA
\item Division of Geological and Planetary Sciences, California Institute of Technology, Pasadena, CA 91125
\item Geological Sciences Department, Stanford University, Stanford, CA 
\item Department of Astronomy, University of Maryland, College Park, MD 20742, USA
\item Space Telescope Science Institute, Baltimore, MD 21218
\item 51 Pegasi b Postdoctoral Fellow
\item NASA Sagan Fellow
\item Department of Physics and Kavli Institute for Astrophysics and Space Research, Massachusetts Institute of Technology, Cambridge, MA 02139
\item Department of Physics and Astronomy, Vanderbilt University, Nashville, TN 37235
\item Department of Aeronautics and Astronautics, Massachusetts Institute of Technology, Cambridge, MA 02139
\end{affiliations}
\vspace{-1cm}
\noindent $^*$\textit{These authors contributed equally to this work}\newline

\begin{abstract}
    Most known terrestrial planets orbit small stars with radii less than 60\% that of the Sun\cite{dressing15b,fulton17}. Theoretical models predict that these planets are more vulnerable to atmospheric loss than their counterparts orbiting Sun-like stars\cite{tarter07, luger15, wordsworth15, shields16}. To determine whether a thick atmosphere has survived on a small planet, one approach is to search for signatures of atmospheric heat redistribution in its thermal phase curve\cite{seager09, selsis11, kreidberg16, koll16}.  Previous phase curve observations of the super-Earth 55 Cancri e (1.9 Earth radii) showed that its peak brightness is offset from the substellar point --- possibly indicative of atmospheric circulation\cite{demory16}.  Here we report a  phase curve measurement for the smaller, cooler planet LHS~3844b, a 1.3\,R$_\oplus$ world in an 11-hour orbit around a small, nearby star.  The observed phase variation is symmetric and has a large amplitude, implying a dayside brightness temperature of 1040$\pm$40~kelvin and a nightside temperature consistent with zero~kelvin (at one standard deviation). Thick atmospheres with surface pressures above 10 bar are ruled out by the data (at three standard deviations), and less-massive atmospheres are unstable to erosion by stellar wind.  The data are well fitted by a bare rock model with a low Bond albedo (lower than $0.2$ at two standard deviations). These results support theoretical predictions that hot terrestrial planets orbiting small stars may not retain substantial atmospheres.  

\end{abstract}

We observed a light curve of the LHS 3844 system with the \Spitzer\ InfraRed Array Camera (IRAC)\cite{fazio04} over 100 hours between UT 4 February 2019 and 8 February 2019 (Program 14204).  We used IRAC's Channel 2 (a photometric bandpass over the wavelength range $4-5\,\mu$m), and read out the $32\times32$ pixel subarray in 2-second exposures. The observations began with a 30-minute dithering sequence to allow the telescope to thermally settle. Following this pre-observation, we employed \Spitzer's Pointing Calibration and Reference Sensor (PCRS) peak-up mode to position the target on the detector's ``sweet spot", a pixel with minimal variation in sensitivity. After the first 60 hours of observation, there was a 3-hour break for data downlink. The data collection recommenced with another 30 minute thermal settling period and continued in PCRS peak-up mode for 40 more hours.  The telescope was re-pointed every 20 hours to keep the image centered on the detector sweet spot. 

We began our analysis with Basic Calibrated Data provided by the Spitzer Science Center (SSC) pipeline, and reduced it with a custom aperture photometry routine\cite{cubillos13}. This routine upsampled each exposure by a factor of 5 in the $X$ and $Y$ dimension and fit a 2D Gaussian profile to determine the image center. We estimated the background from the median value in an annulus 7 to 15 pixels from the target center. Bad pixels were identified and masked based on iterative $\sigma$-clipping over groups of 64 exposures.  We then summed the flux in a fixed aperture centered on the target. We varied the aperture size from 2 to 4 pixels in 0.5 pixel increments, and selected a 2.5-pixel aperture to minimize noise in the resulting light curve.

We fit the extracted light curve with a simultaneous model of the astrophysical signal and the instrument behavior. The astrophysical signal consisted of a transit model and a first-degree spherical harmonics temperature map to represent the planet's thermal phase variation \cite{kreidberg15a, louden18}. The instrument model had two components: a two-dimensional spline fit to Spitzer's pixel sensitivity variations and a linear scaling with the half-width of the point spread function in both the $X$ and $Y$ direction\cite{stevenson12, lanotte14}.  We determined the best-fit model parameters with a least squares minimization routine and estimated uncertainties with differential evolution Markov Chain Monte Carlo (MCMC). We explored  alternative models for both the instrument systematics and the planet's thermal phase variation, and obtained consistent results with our nominal model (see Methods for further detail).

\begin{figure}[t!]                                  
\includegraphics[width=0.93\textwidth]{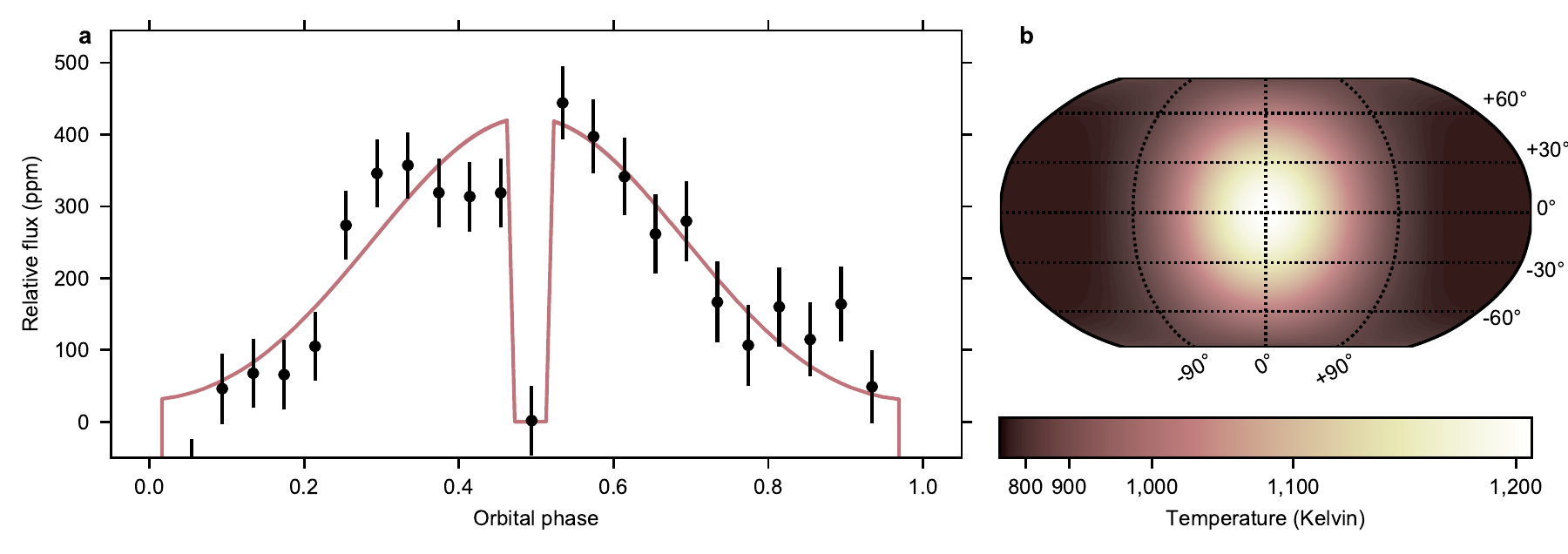}
\caption{\textbf{The 4.5\,$\mu$m thermal phase curve of LHS~3844b and best fit temperature map.} \textbf{a,}  Planet-to-star flux binned over 25 equally spaced intervals over the planet's 11.1-hour orbital period (points with $1\,\sigma$ uncertainties) compared to the best fit phase curve (line). The data are normalized such that the relative flux is zero when the planet is eclipsed by the star at orbital phase 0.5. \textbf{b,} The spherical harmonic temperature map used to generate the phase variation model. The planet's substellar point corresponds to the latitude and longitude ($0^\circ$, $0^\circ$). We note that the spherical harmonics model includes north-south temperature variation, but only east-west variation is constrained by the data.}
\label{fig:phasecurve}  
\end{figure}

Figure 1 shows the measured thermal phase curve and best fit temperature map. The secondary eclipse depth is $380 \pm 40$ ppm and the peak-to-trough amplitude of the phase variation is $350 \pm 40$ ppm. The values correspond to a dayside brightness temperature of $1040 \pm 40$~K, and a nightside brightness temperature broadly consistent with zero ($0-710$~K at $1\sigma$ confidence). The longitude of peak brightness is consistent with zero degrees ($-6\pm6^\circ$).  The inferred planet-to-star radius ratio is $R_p/R_s = 0.0641 \pm 0.0003$, which is consistent with the optical light radius ratio\cite{vanderspek19}. The secondary eclipse time agrees with expectations for a zero eccentricity orbit. We find no evidence for transit time variations (the transit times deviate from a linear ephemeris at $0.1 \sigma$ confidence). From a joint fit to the \Spitzer\ and \TESS\ transit times, we revise the time of central transit to $2458325.22559\pm0.00025\,\mathrm{BJD_{TDB}}$ and the orbital period to $0.4629279 \pm 0.0000006$ days.  We also established upper limits for transits of other planets in this system, based on a joint fit to the \TESS\ and \Spitzer\ data.  In the orbital period range from 0.5 to 6 days, our $3\sigma$ upper limit corresponds to $0.6 R_\oplus$, and from 6 to 12 days our $3\sigma$ upper limit is $0.8R_\oplus$.  

\begin{figure}[t!]
\begin{center}
\includegraphics[width=0.8\textwidth]{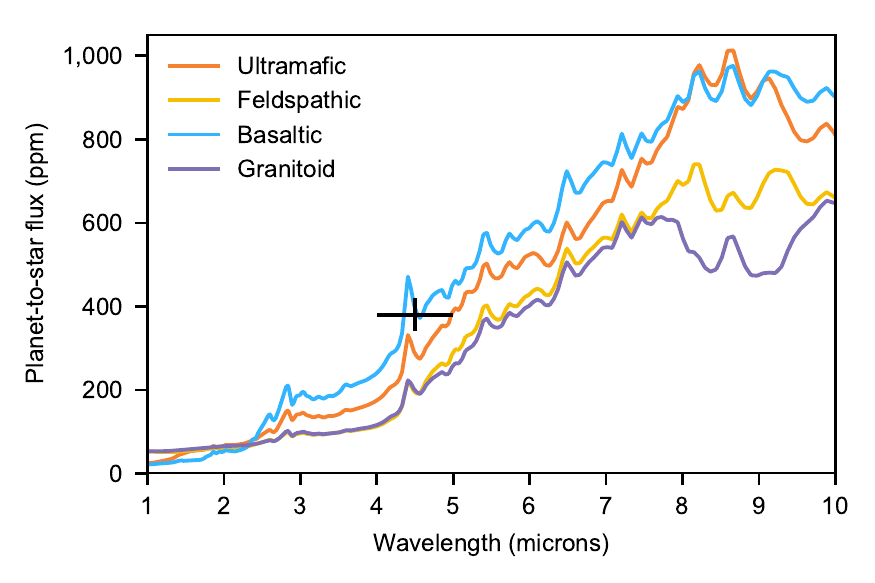}
\end{center}
    \caption{\textbf{Predicted emission spectrum for a range of surface compositions.} The measured planet-to-star flux (shown in black) is most consistent with a pure dark basaltic surface. Mixed surface compositions require at least 40\% basalt or 75\% ultramafic rock to be consistent with the data at $3\sigma$ confidence.} 
\label{fig:surfaces}  
\end{figure}

The phase curve is consistent with expectations for a synchronously rotating bare rock --- a completely absorptive surface in instantaneous thermal equilibrium that radiates isotropically (shown in Extended Data Figure 1). In this simple picture, the amplitude of the phase curve requires the surface to be very absorptive, with an upper limit on the Bond albedo of 0.2 at 2$\sigma$ confidence.   We modeled the emission spectra of several rocky surfaces\cite{hu2012surface} and compared with the measured planet-to-star flux (shown in Figure 2). We considered multiple geologically plausible planetary surface types, including primary crusts that form from solidification of a magma ocean (ultramafic and feldspathic), secondary crust that forms from volcanic eruptions (basaltic), and a tertiary crust that forms from tectonic re-processing (granitoid).  Governed by the reflectivity in the visible and the near-infrared and the emissivity in the mid-infrared, the surface types have distinct emission spectra. The measured planet-to-star flux for LHS~3844b is most consistent with a basaltic composition.  Such a surface is comparable to the lunar mare and Mercury, and could result from widespread extrusive volcanism\cite{depater01}. Pure feldspathic and granitoid compositions provide a poor fit to the data, and must be mixed with at least 40\% basaltic or 75\% ultramafic surface to to be consistent with the measured eclipse depth at $3\sigma$.

\begin{figure}[t!]     
\begin{center}
    \includegraphics[width=0.8\textwidth]{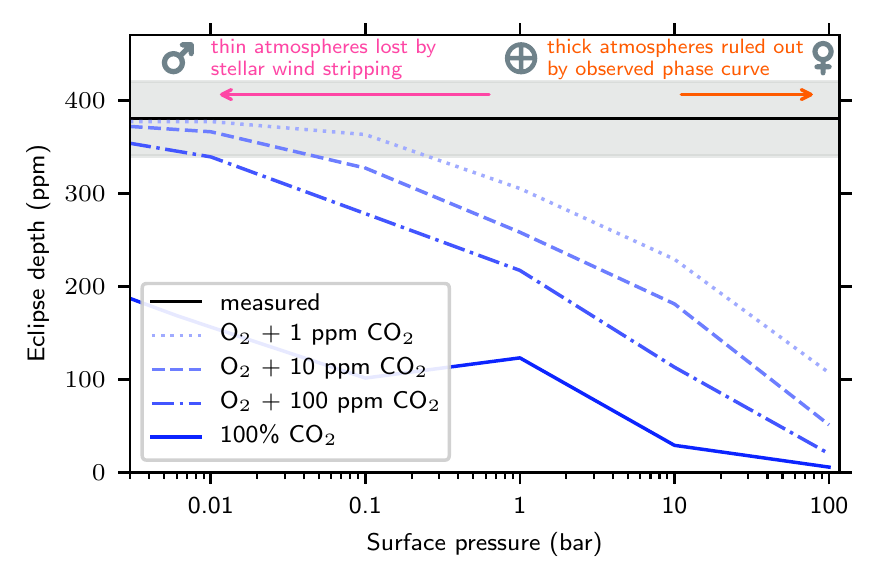}
\end{center}
    \caption{\textbf{Predicted $4.5\,\mu$m eclipse depths for model atmospheres compared to the measured value.} The gray region represents the $1\sigma$ uncertainty on the measured eclipse depth. The model atmospheres are composed of oxygen/carbon dioxide mixtures and account for heat transport to the planet's nightside. We indicate the surface pressures of Mars, Earth, and Venus with gray symbols. The orange arrow indicates surface pressures that are inconsistent with the observed eclipse depth ($> 3\sigma$ confidence); the pink arrow shows surfaces pressures that are unstable to erosion by stellar wind.}
\label{fig:modelatmo}  
\end{figure}

We also explored the possibility that the planet has an atmosphere. We developed a simple model to account for both the atmospheric heat redistribution and absorption features from plausible chemical compositions.  We parameterized the day-night atmospheric heat redistribution with a scaling that is based on analytic theory\cite{koll16} and that accounts for the dynamical effects of surface pressure, $p_s$ and atmospheric optical thickness, $\tau_{LW}$. See Methods for the equation and a validation of the scaling against dynamical models. To estimate the planet's eclipse depth in the \Spitzer\ 4.5\,$\mu$m bandpass, we calculated 1D radiative transfer models\cite{morley17} tuned to match the heat redistribution scaling. Motivated by atmospheric evolution models of hot, terrestrial planets\cite{wordsworth13,wordsworth14,schaefer16}, we considered model atmospheric compositions that are mixtures of oxygen (O$_2$) and carbon dioxide (CO$_2$), over a range of surface pressures from $0.001 - 100$ bar. We also considered nitrogen (N$_2$) mixtures with trace carbon dioxide.

Figure 3 shows the predicted eclipse depths for O$_2$/CO$_2$ models compared to the measured values.  Higher surface pressures correspond to smaller eclipse depths (implying a cooler dayside) because thick atmospheres are more efficient at transporting energy to the planet's nightside. Higher CO$_2$ abundances also decrease the predicted eclipse depths, due to strong absorption by CO$_2$ in the \Spitzer\ $4.5\,\mu$m bandpass that pushes the photosphere to higher, cooler layers.  Overall, we find that the best fit models have surface pressures below 0.1 bar. Carbon dioxide-dominated atmospheres are ruled out for surface pressures as low as Mars (0.006 bar), and surface pressures above 10 bar are ruled out for all compositions we consider (greater than $3\sigma$ confidence). For N$_2$ mixtures, high surface pressures were ruled out at even higher confidence (e.g., a 10 bar N$_2$ atmosphere with 1 ppm CO$_2$ is excluded at 6$\sigma$).

As an independent test of these results, we also fit the measured phase curve variation with an energy-balance model that computes reflection, longitudinal heat redistribution, and thermal emission. The model is parameterized by the planet's Bond albedo, the ratio of radiative to advective timescales, and a greenhouse warming factor\cite{hu15,angelo2017}.  For photometric data like ours, the Bond albedo and greenhouse factor are degenerate, but we included both parameters to capture the possibility that the observed brightness temperature is different from the energy balance temperature.  Based on an MCMC analysis, we infer a value of $\tau_\mathrm{rad}/\tau_\mathrm{adv} < 0.3$ at $2\sigma$ confidence. For a typical wind speed of 300 m s$^{-1}$ for high mean molecular weight atmospheres\cite{Kataria14,Zhang17}, and assuming a surface gravity of 16 m s$^{-2}$ consistent with an Earth-like bulk composition (the planet's mass is not yet known), this requirement implies a photospheric pressure less than 0.06 bar, in agreement with our finding that the data are fit well by tenuous O$_2$ and N$_2$-dominated atmospheres with trace amounts of CO$_2$.

\begin{figure}[t!]
\begin{center}
\label{fig:atmoevol}  
    \includegraphics[width=0.8\textwidth]{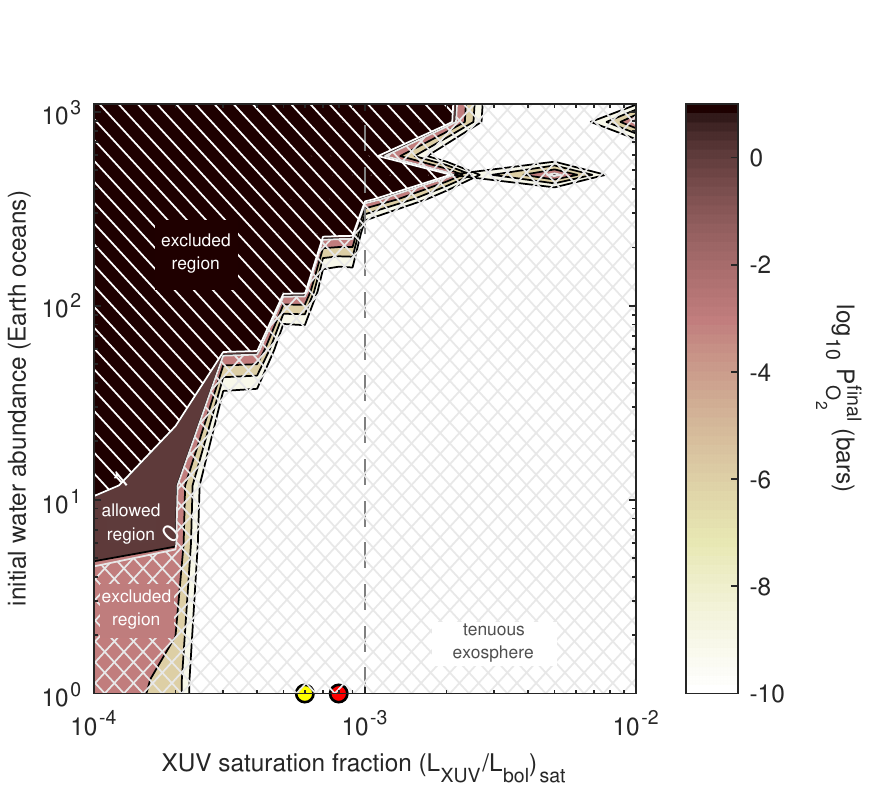}
    \caption{\textbf{Final atmospheric pressure after 5 Gyr of energy-limited atmospheric escape.} The contours indicate final surface pressure as a function of the XUV saturation fraction and the initial water abundance.  All water is lost from all cases considered here by the end of the simulation, but residual O$_2$ gas remains. Present-day surface pressures greater than 10 bar (single hatched region) are excluded based on the observed phase curve amplitude, and pressures below 0.7 bar are unstable to erosion by stellar wind (cross hatched region). The vertical dashed line marks the nominal saturation fraction for LHS 3844.  The yellow and red dots on the x-axis show the  $L_\mathrm{XUV}/L_\mathrm{bol}$ of the early Sun and a young M-dwarf (AD Leo), respectively\cite{ribas05,chadney15}. }
\end{center}
\end{figure}

To explore whether a tenuous atmosphere or no atmosphere at all is more likely for LHS 3844b, we modeled atmospheric escape over the planet's lifetime\cite{schaefer16}.  The initial atmosphere is assumed to be pure water, which can either (1) dissolve within a magma ocean formed by accretion-induced heating during planet formation, or (2) photolyze into hydrogen and oxygen due to high energy stellar radiation in the XUV wavelength range ($0.1 - 100$ nm)\cite{ribas05}. Most of the atomic hydrogen and oxygen escape to space, but some remnant oxygen reacts with the magma ocean or remains in the atmosphere as O$_2$. We assume that the early XUV flux is a constant fraction of the stellar bolometric luminosity (the ``saturation fraction") until 1 Gyr and then decays with time following a power law \cite{ribas05,luger15}. In our model, we vary both the initial planetary water abundance and the XUV saturation fraction. Figure 4 shows the resulting surface pressures compared to the $3\sigma$ upper limit of 10 bar obtained in this work. For a typical saturation fraction for low-mass stars\cite{chadney15}, we find that the initial planetary water abundance could not exceed 240 Earth oceans. For lower initial water abundances or higher XUV saturation, the atmosphere is entirely lost. Further, we estimate that $1 - 10$ bars of atmosphere could be eroded by stellar winds (see Methods for details).  Given that thick atmospheres are ruled out by the data and thin atmospheres are unstable over the planet's lifetime, LHS 3844b is most likely a bare rock, unless a thin atmosphere is continually replenished over time.

The results presented here motivate similar studies for less-irradiated planets orbiting small stars.  Cooler planets are less susceptible to atmospheric escape and erosion, and may provide a friendlier environment for the evolution of life. In coming years this hypothesis can be tested, thanks to the infrared wavelength coverage of the \emph{James Webb Space Telescope} and the influx of planet detections expected from current and future surveys.

\noindent\textbf{Data availability}\\
The raw data used in this study is publicly available at the \Spitzer\ Heritage Archive, \url{https://sha.ipac.caltech.edu/applications/Spitzer/SHA}. 

\noindent{\textbf{Code availability}\\
    We processed and fit the data with the open-source POET pipeline, available at \url{https://github.com/kevin218/POET}. We used the code version corresponding to commit ID \texttt{adbe62e7b733df9541231e8d1e5d32b7e2cdad76}.\\

\subsection{\large{References}}~\\
 \newcommand{\noop}[1]{}

\begin{addendum}
 \item D.D.B.K. was supported by a James S. McDonnell Foundation postdoctoral fellowship. R.H. is supported in part by NASA Grant No. 80NM0018F0612. The research was carried out at the Jet Propulsion Laboratory, California Institute of Technology, under a contract with the National Aeronautics and Space Administration. A.V.'s work was performed under contract with the California Institute of Technology (Caltech)/Jet Propulsion Laboratory (JPL) funded by NASA through the Sagan Fellowship Program executed by the NASA Exoplanet Science Institute. D.C. acknowledges support  from the John Templeton Foundation. The opinions expressed in this publication are those of the authors and do not necessarily reflect the views of the John Templeton Foundation.

 \subsection{Contributions} L.K. conceived the project, planned the observations, and carried out the primary data reduction. D.D.B.K., C.M., and R.H. contributed equally to this work. They ran theoretical models for the planet's atmosphere and surface. L.S. provided atmospheric evolution models. D.D., K.B.S., J.D., A.V., D.B., and X.G. contributed to the data analysis. K.S. modeled the stellar spectrum. I.C., D.C., D.W.L., A.L., G.R., S.S., and R.V. provided useful comments on the manuscript and assisted with the observing proposal.                                                 

 \item[Competing Interests] The authors declare that they have no
competing financial interests.
 \item[Correspondence] Correspondence and requests for materials
should be addressed to L.R.K.~(email: laura.kreidberg@cfa.harvard.edu).
\end{addendum}

\begin{methods}

\subsection{Data analysis}
In this section we provide additional description of the data analysis. Our nominal model for the planet's phase variation was a first-degree spherical harmonics temperature map for the planet. For each point in the time series, we computed the planet-to-star flux ratio for the viewing geometry at that time\cite{louden18}. We multiplied the planet-to-star flux by a transit model to account for the drop in stellar flux when the planet obscures the star.

We used a differential evolution Markov Chain Monte Carlo (deMCMC) algorithm to obtain the posterior distribution of the model parameters\cite{braak06}. The free parameters for the fit were the planet-to-star radius ratio, the time of central transit, a linear limb-darkening parameter, and four coefficients for the spherical harmonics model.  The fit also had free parameters for a normalization constant and linear scaling of the PRF width in both the $X$ and $Y$ direction\cite{lanotte14}. We fixed the other orbital parameters on the best fit values from the discovery paper. We initially allowed the secondary eclipse time to vary, but found that it was consistent with expectations for zero eccentricity, so our final fit assumed a circular orbit. We also applied a prior that assigned zero probability to solutions with temperature maps that dropped below zero.

\begin{figure}[t!]
\makeatletter
\renewcommand{\fnum@figure}{Extended Data Figure 1}
\makeatother
    \begin{center}
\includegraphics[width=0.8\textwidth]{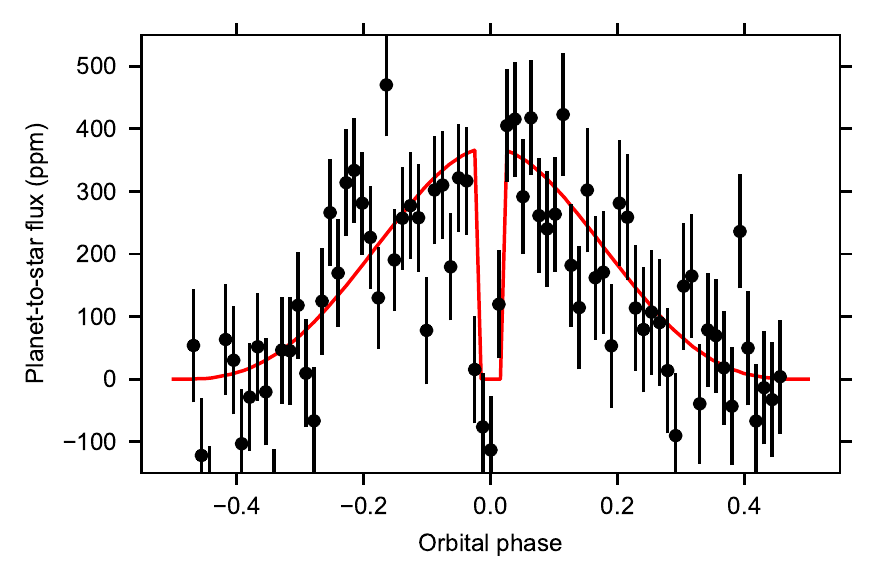}       
    \end{center}
\caption{\textbf{The LHS~3844b phase curve compared to predictions for a bare rock.} The model assumes the incident stellar flux is totally absorbed (zero Bond albedo), and is reradiated instantaneously and isotropically. The error bars correspond to $1\sigma$ uncertainties.}
\label{fig:rock}                                              
\end{figure} 

Over the observation, the telescope was repointed several times (see Extended Data Figure 2). We assumed that the astrophysical parameters were the same for all five pointings, but allowed the systematics parameters to vary. We fit all the data simultaneously. We masked two short segments of data that showed correlated noise in the residuals (marked in gray in Extended Data Figure 2). As a test of the fit quality, we made an Allan deviation plot (Extended Data Figure 3). The rms bins down as the square root of the number of data points per bin, as expected for photon noise-limited statistics. We rescaled the estimated uncertainty per data point by a factor of 1.07 to achieve a reduced $\chi^2$ value of unity for the best fit. We then ran the deMCMC with four chains until the Gelman Rubin statistic dropped below 1.01. 

\begin{figure}[t!]
\makeatletter
\renewcommand{\fnum@figure}{Extended Data Figure 2}
\makeatother
    \begin{center}
\includegraphics[width=0.93\textwidth]{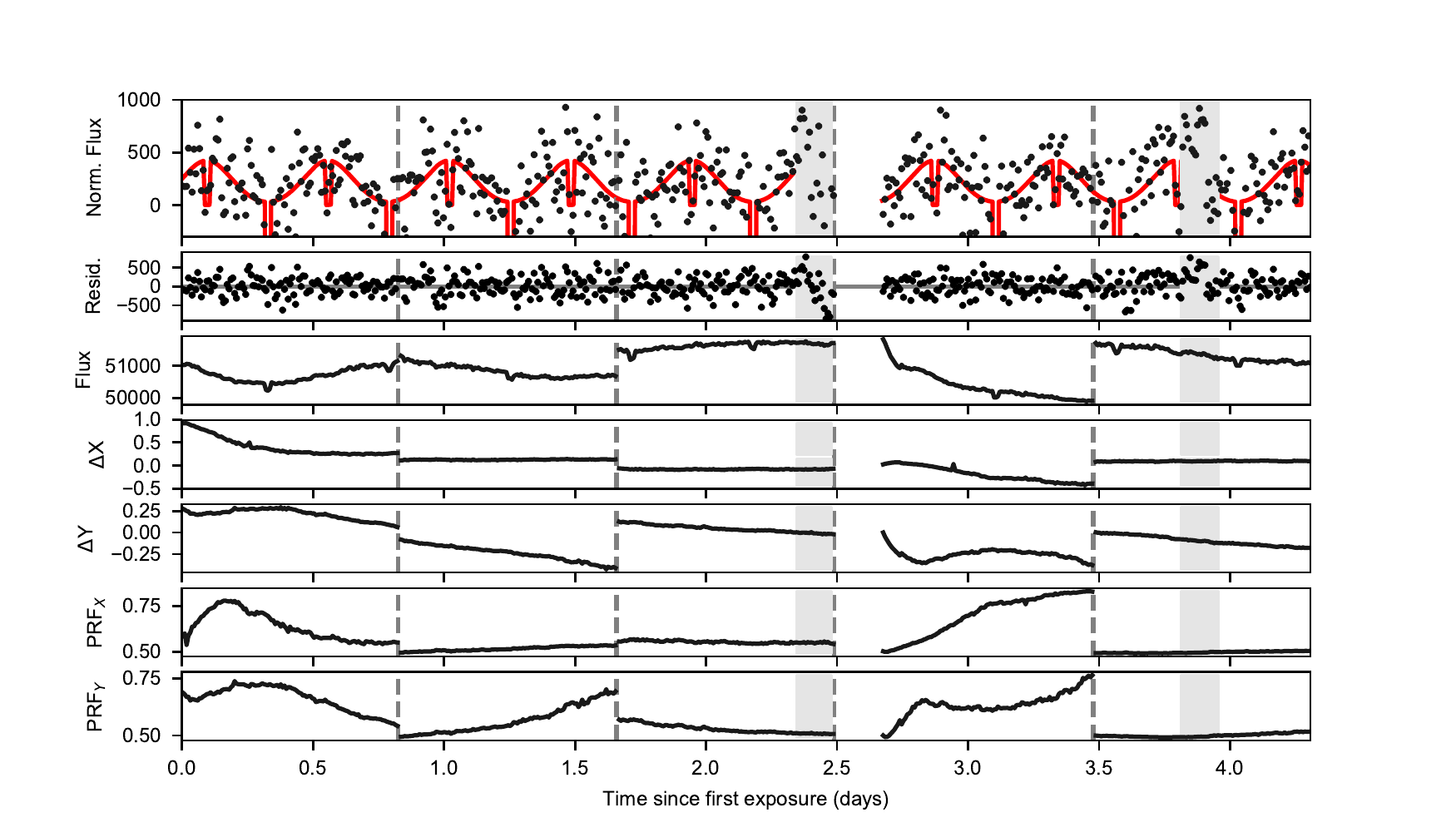}       
    \end{center}
\caption{\textbf{LHS~3844b light curve and diagnostics.}  The top panel shows the normalized light curve (points) and best fit model (red line). The data are binned in 10-minute increments. Regions marked in gray were masked in the fit due to time-correlated noise in the residuals. The second panel shows the residuals to the best fit model. The recorded flux in $\mu$Jy is given in the third panel. Panels four and five show the shift in $X$ and $Y$ of the target centroid on the detector. The final two panels show the width of the PRF in the $X$ and $Y$ direction. Changes in telescope pointing are marked with dashed vertical lines. The gap after 2.5 days is due to data downlink.}
\label{fig:diagnostics}                                              
\end{figure} 

In addition to the spherical harmonics model, we also tested a sinusoid model, which has been commonly used to fit other phase curve data\cite{demory16}. The only reason the two would produce different phase curves is that the transformation from temperature to flux (the Planck function) is not linear.  The sinusoid peak-to-trough amplitude was $402\pm46$ ppm, the eclipse depth as $425\pm54$ ppm, and the longitude of peak brightness is consistent with the substellar point ($-3.0\pm6.0$ degrees). The phase curve amplitude and eclipse depth are consistent within $1\sigma$ with the results from the spherical harmonics model. We inverted the sinusoid to estimate the planet's temperature as a function of longitude\cite{cowan08} and found that the nightside temperature was below zero for the best fit sinusoid. Similar behavior has been observed for other phase curves with large amplitudes, and in general can be remedied by the addition of odd harmonics in the sinusoid model \cite{keating17}. In this work, we focus on the spherical harmonics fit where we impose positive nightside temperatures.

\begin{figure}[t!]
\makeatletter
\renewcommand{\fnum@figure}{Extended Data Figure 3}
\makeatother
    \begin{center}
\includegraphics[width=0.8\textwidth]{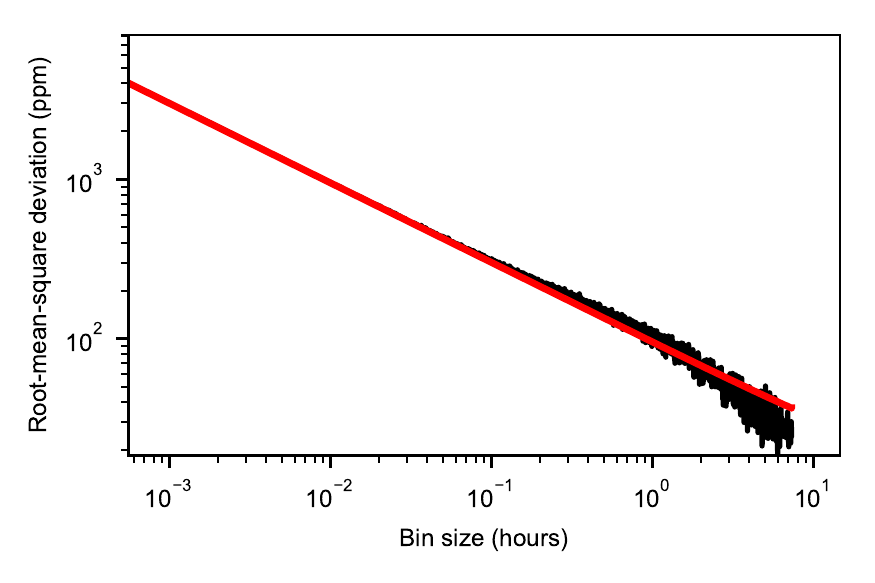}       
    \end{center}
\caption{\textbf{Allan deviation plot for the residuals from the full phase curve fit.}  The data bin down as expected with the photon noise on timescales from 2 s (1 exposure) to 5.5 hours ($10^4$ exposures).}
\label{fig:allen}                                              
\end{figure}

\subsection{Effects of red noise}
    For the main analysis reported in this paper, we trimmed small sections of data that exhibited time-correlated noise (illustrated in Extended Data Figure 2). As a test, we also performed a full MCMC fit to the entire data set, accounting for red noise noise on the timescale of the eclipse duration\cite{pont06}.  For the full data set, we obtained an eclipse depth and sine curve amplitude of $490\pm70$ and $450\pm70$ ppm. These results are consistent with the values reported for the trimmed data and do not change the conclusions of the paper.  

\subsection{Independent data reduction and analysis}
We checked our results using an independent analysis using first-order Pixel Level Decorrelation (PLD)\cite{deming15}.  First-order PLD has limited applications to phase curves because the image motion can exceed the limits of the first order approximation.  Therefore our PLD analysis focused on checking the amplitudes of the transit and secondary eclipse.  We fit eclipse and transit separately, each using a range in orbital phase that is as large as possible (adjusted by trial and error), while still preserving the highest possible precision in the fits.  We reduced red noise by fitting to binned data\cite{deming15}. Our PLD code selected the optimum bin size and the optimum radius of the photometric aperture, based on minimizing the chi-squared value in a fit to the Allan deviation curve\cite{garhart19}.  For the eclipse, we binned the data by 392 points, versus 189 point binning for the transit.  We binned the modeled eclipse and transit curves by the same amounts as the data in order to avoid bias in the fits, and we verified that the derived transit and eclipse amplitudes do not vary systematically with the bin size.  

\begin{figure}[t!]                                              
\makeatletter
\renewcommand{\fnum@figure}{Extended Data Figure 4}
\makeatother
    \begin{center}
\includegraphics[width=0.6\textwidth]{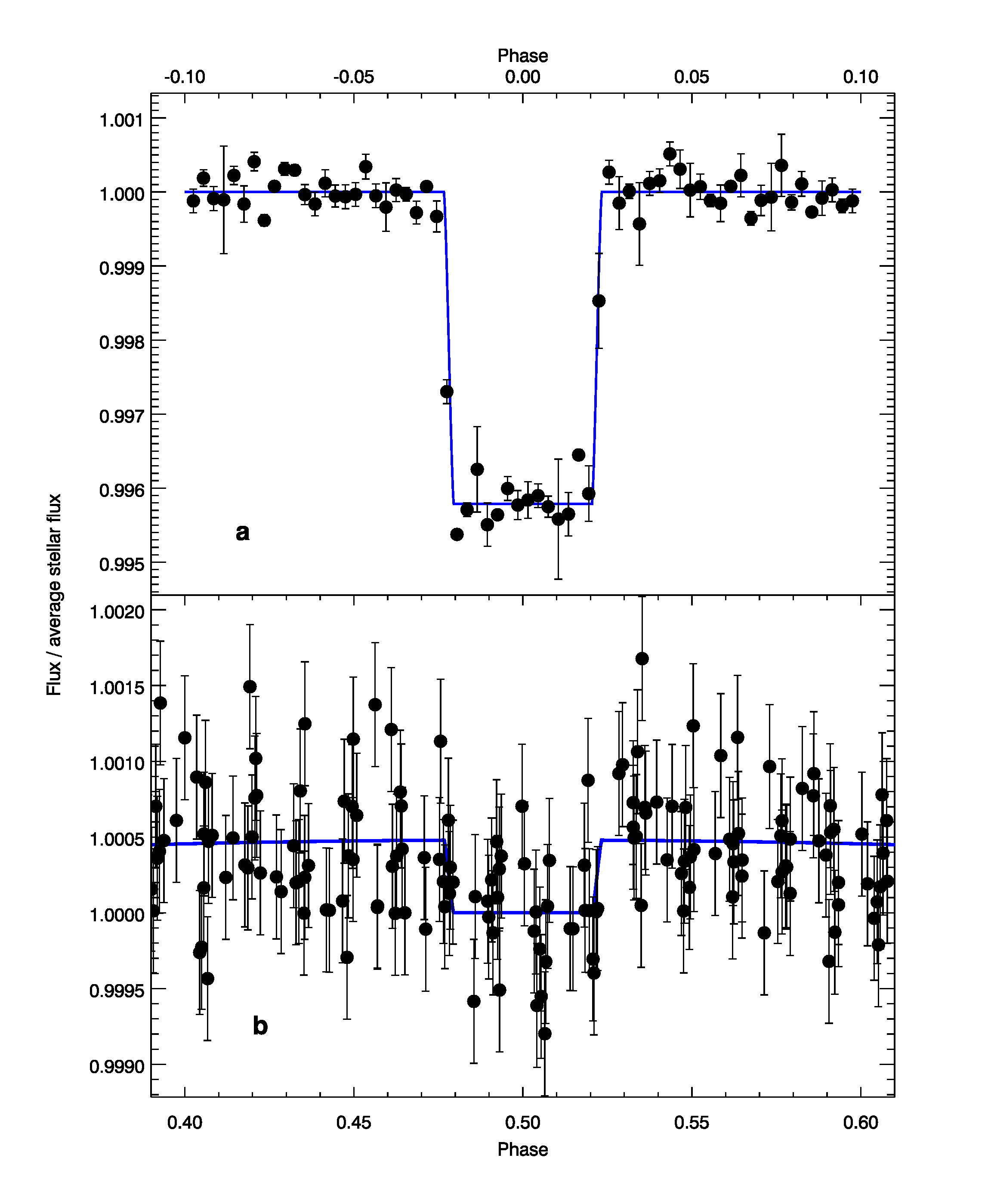}       
    \end{center}
    \caption{\textbf{Independent transit and eclipse fits.} \textbf{a,} Spitzer transit of LHS~3844b at 4.5\,$\mu$m, decorrelated using PLD (see text), and phased and binned for clarity of illustration.  \textbf{b,} Spitzer secondary eclipse of LHS~3844b at 4.5\,$\mu$m, decorrelated in a manner similar to the transit, but using a different binning. (The binning used in this figure is merely for illustration, and is not the same as the binning used by our PLD code.) The error bars correspond to 1$\sigma$ uncertainties.}
\label{fig:2panel}                                              
\end{figure} 
Since the orbital ephemeris has been determined recently and precisely from TESS observations\cite{vanderspek19}, we fixed the phase of transit to 0.0 and the phase of secondary eclipse to 0.5.  We used quadratic limb darkening for the transit\cite{claret12}.  Since infrared limb darkening is small, we also fixed those coefficients during the transit fitting process. The PLD code determined the best fit transit and eclipse amplitudes using multi-variate linear regression, and then estimated the uncertainties with MCMC. The independent variables in the fit were the 12 basis pixels\cite{garhart19}, a linear ramp in time, and the modeled transit/eclipse curve.  The MCMC consisted of a 50,000-step burn-in segment, followed by a 500,000-step chain using Metropolis-Hastings sampling.  We ran independent chains to verify convergence.  The posterior distributions for eclipse and transit amplitude are closely Gaussian, and centered on the best-fit values from the multi-variate regression.  Because the derived amplitudes vary stochastically with different bin sizes, we include that effect in the error bars on the amplitudes, by repeating the regressions over a range of bin sizes that give acceptable fits.  We add the dispersion in amplitudes from that process in quadrature with the uncertainty from the MCMCs to arrive at total error bars for the results (the MCMC uncertainties dominate).

Our best fits for the transit and eclipse are illustrated in Extended Data Figure 4.  Our secondary eclipse amplitude from the PLD fits is $439\pm54$ ppm, differing from our value from the nominal value from the phase curve analysis ($380\pm40$) by $0.9\sigma$.  The brightness of the star in the infrared and the small limb darkening facilitate a precise radius measurement  from the \Spitzer\ transits.  We find $R_p/R_s = 0.0641\pm0.0003$, in excellent agreement with the optical wavelength radius, $R_p/R_s = 0.0635 \pm 0.0009$\cite{vanderspek19}.   

As a test, we also attempted a PLD fit to the entire phase curve. For this analysis, we modified the PLD technique by adding quadratic terms for each pixel while neglecting crossterms\cite{Zhang17}. We selected a 5$\times$5 pixel aperture centered on LHS~3844 from which to extract our light curve. In addition to the coefficients for each pixel, we also adopted a quadratic function in time to correct for the long term drift in photometry. Each AOR was reduced independently. Prior to fitting the data, we rejected all data points more than 10 median absolute deviations away from a 1 hour wide sliding median in order to eliminate the effect of large outliers in our data. Prior to fitting our phase curve, we eliminate all data points that occur during the primary transit of LHS~3844b and data points that occur within 1 transit duration of transit to avoid potentially biasing our result. We found that the PLD reduction produces consistent results with our primary analysis: the best fit sinusoidal model had a peak-to-trough phase curve amplitude of 310 ppm and a secondary eclipse depth of 335 ppm. However, the fit has time-correlated noise that makes it challenging to robustly estimate uncertainties. For these data, the target centroid motion is large enough that the assumptions behind PLD break down, and the results from the 2D-spline mapping technique\cite{stevenson12} are more robust.

\begin{figure}[t!]
\makeatletter
\renewcommand{\fnum@figure}{Extended Data Figure 5}
\makeatother
    \begin{center}
        \includegraphics[width=0.8\textwidth]{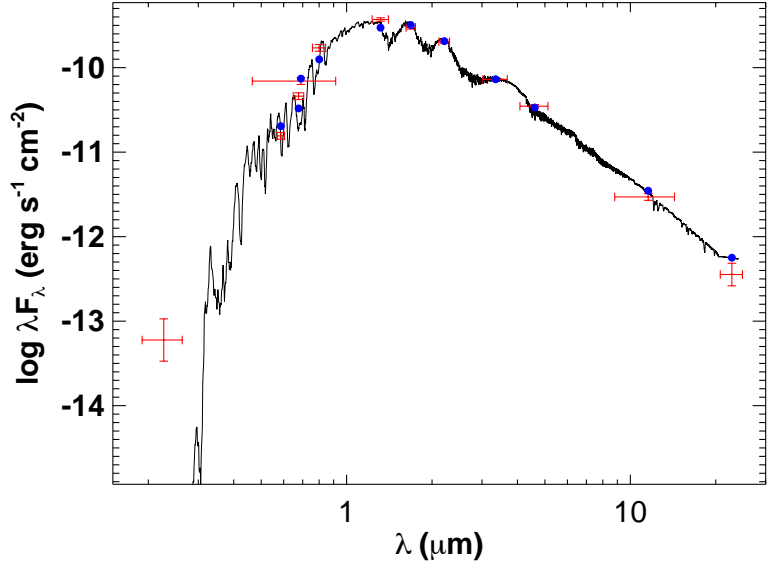}       
    \end{center}
\caption{\textbf{The best fit spectral energy distribution for LHS~3844.} The  vertical error bars represent $1\sigma$ uncertainties. The horizontal error bars represent photometric bandpasses.}
\label{fig:sed}                                              
\end{figure} 

\subsection{SED Fit}
The theoretical models in this paper require an estimate of the stellar spectrum. We estimated the spectrum with a spectral energy distribution (SED) fit using all available photometry. The SED fit (shown in Extended Data Figure 5) gave a metallicity of $\mathrm{[Fe/H]} = 0.0^{+0.0}_{-1.0}$ and a bolometric luminosity $F_\mathrm{bol} = 3.52\pm0.33\,\mathrm{erg/s/cm}^2$ (measured at Earth). From these values and the Gaia parallax\cite{gaiadr2}, we derived a stellar radius $R_* = 0.178 \pm 0.012$ (assuming a stellar effective temperature $T_\mathrm{eff} = 3036\pm77$~K)\cite{vanderspek19} and a stellar mass $M_* = 0.158\pm0.004\,M_\odot$\cite{mann19}. There is  chromospheric emission in the GALEX FUV band, which could contribute to present-day photoevaporation of the planetary atmosphere.

\subsection{Model for atmospheric heat redistribution}
We computed the planet's broadband dayside-averaged brightness temperature using the following scaling:
\begin{eqnarray}
  T_{day} & = & T_* \sqrt{\frac{R_*}{d}} (1-\alpha_B)^{1/4} f^{1/4}
\end{eqnarray}
where
\begin{eqnarray}
 f & = & \frac{2}{3} - \frac{5}{12} \times
                        \frac{\tau_{LW} \left(\frac{p_s}{1
                        \mathrm{bar}}\right)^{2/3}\left(\frac{T_{eq}}{600\mathrm{K}}\right)^{-4/3}}{2
                        +\tau_{LW} \left(\frac{p_s}{1
                        \mathrm{bar}}\right)^{2/3}\left(\frac{T_{eq}}{600\mathrm{K}}\right)^{-4/3}}.
\end{eqnarray}
Here $T_*$ is the stellar effective temperature, $R_*$ is the stellar radius, $d$ is the semi-major axis, $\alpha_B$ is the planetary albedo, and $T_{eq}$ is the planet's equilibrium temperature. The derivation for the scaling will be presented elsewhere (Koll et al., submitted). To validate the scaling, Extended Data Figure 6 compares the dayside eclipse depths predicted by the scaling against the dayside eclipses simulated with a general circulation model (GCM) with tidally-locked orbital parameters and semi-grey radiative transfer. The scaling successfully captures the main variation in the GCM's day-night heat redistribution. 

To include our scaling in the 1D radiative transfer model, we calculated the 
broadband optical thickness $\tau_{LW}$ for a given atmospheric composition 
and surface pressure based on the atmosphere's attenuation of the surface's thermal emission,
\begin{eqnarray}
  \tau_{LW} & = & -\ln\left[  \frac{\int e^{-\tau_\lambda}
                  B_\lambda(T_s)d\lambda}{\int
                  B_\lambda(T_s)d\lambda} \right].
\end{eqnarray}
Here $B_\lambda$ is the Planck function, $T_s$ is the surface temperature, and $\tau_\lambda$ is the atmosphere's column-integrated optical thickness at a given wavelength computed with the 1D radiative transfer model.

\begin{figure}[t!]                                          
\makeatletter
\renewcommand{\fnum@figure}{Extended Data Figure 6}
\makeatother
    \begin{center}
    \includegraphics[width=0.8\textwidth]{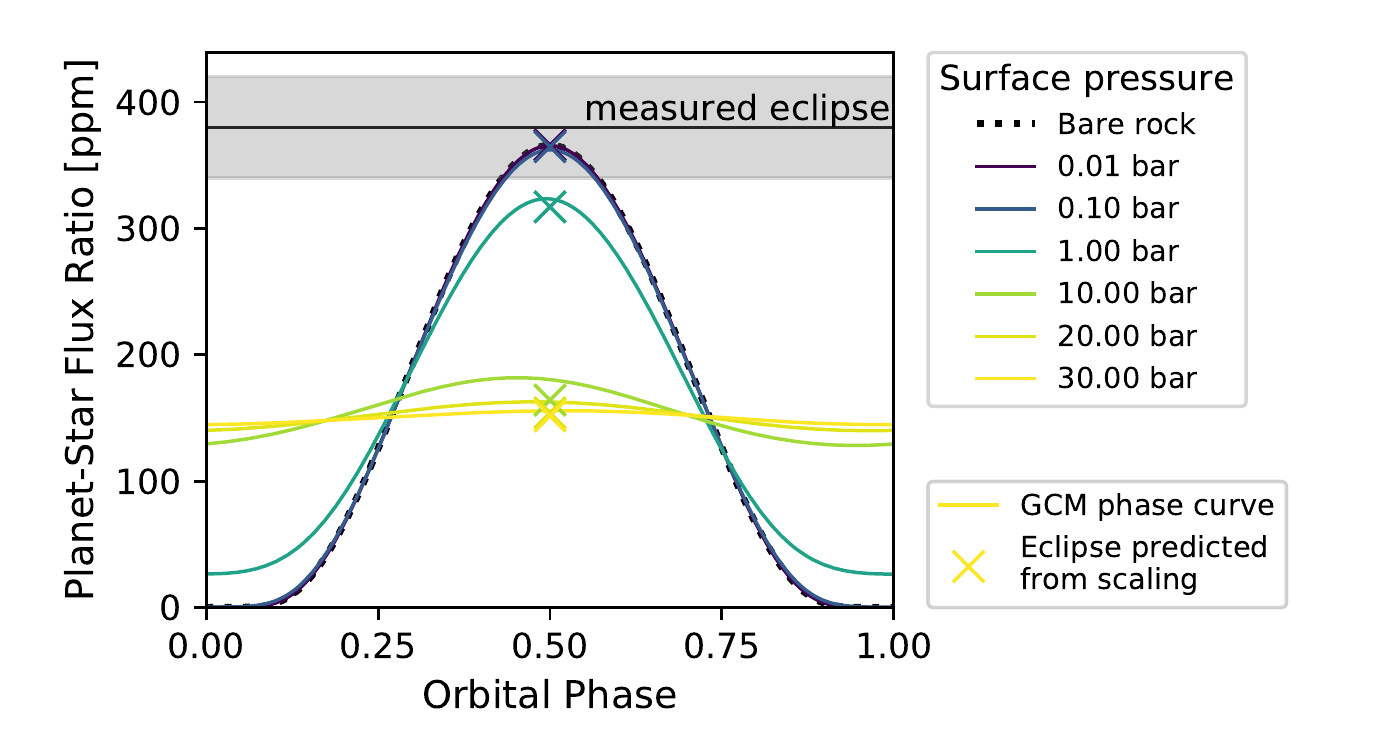}
    \end{center}
    \caption{\textbf{Thermal phase curve of LHS~3844b in the Spitzer bandpass as a function of surface pressure, simulated with the FMS general circulation model\cite{frierson2006,merlis2010,koll15}.} We used semi-grey radiative transfer and increased the longwave optical thickness linearly with surface pressure, $\tau_{LW}=p_s/\mathrm{1 bar}$. The grey region shows the measured secondary eclipse with 1$\sigma$ uncertainty, colored lines show simulated phase curves, and crosses show the dayside eclipses predicted by our analytic scaling. Thin atmospheres with less than 1 bar surface pressure are indistinguishable from a bare rock, while thick atmospheres become increasingly uniform.}
\label{fig:phasecurves}                                              
\end{figure} 

\subsection{Atmospheric escape due to stellar wind}
Interaction with a stellar wind can also be a significant source of atmospheric erosion. We estimated the erosion for LHS~3844b based on ion escape rates calculated for Proxima Centauri b of $10^{26} - 10^{27}\, \mathrm{s}^{-1}$ (equivalent to $3-30$ kg/s of mass loss)\cite{dong17}. Proxima Centauri is approximately the same stellar type as LHS 3844. Scaling for orbital distance, the stellar wind flux onto LHS~3844b is $\sim10\times$ larger than  for Proxima~b, corresponding to a mass loss rate of $30-300$ kg/s. Assuming a constant stellar wind flux over the planet's lifetime, this implies a total mass loss of $10^{18}-10^{19}$ kg, equivalent to $0.7 - 7$ bars. The stellar wind flux was likely higher during the star's active period, so this value is a lower limit to the amount of escape that could be driven by the stellar wind alone.

\subsection{Stability to atmospheric collapse}
We explored the possibility of atmospheric collapse by comparing LHS 3844b to previously published models of synchronously rotating planets with CO$_2$-dominated atmospheres\cite{wordsworth15}.  In the simulations, planets with three times Earth insolation ($S_\oplus$) were stable to collapse for surface pressures of 0.03 bar (assuming $M_p = 1\,M_\oplus$) or 0.4 bar ($10\,M_\oplus$).  The higher the insolation, the more stable the atmosphere is to collapse. Since LHS 3844b is much more highly irradiated than the range of planets considered ($70\,S_\oplus$), we expect surface pressures below $\sim0.1$ bar are stable. For the non-synchronous case, the planet is more evenly heated, making atmospheric collapse even less likely.

\end{methods}

\end{document}